\documentclass[12pt]{iopart}

\usepackage{iopams}
\usepackage{braket}
\usepackage{amsfonts}
\usepackage{color}
\usepackage[dvipsnames]{xcolor}
\usepackage{amssymb}
\usepackage{mathrsfs}
\usepackage{graphicx}
\usepackage{lmodern}
\usepackage{lineno}
\usepackage{hyperref}
\usepackage[normalem]{ulem}

\newcommand{\beq}{\begin{eqnarray}}
\newcommand{\eeq}{\end{eqnarray}}

\def\keywords#1{\vspace{10pt}
     \begin{indented}
     \item[]\rm Keywords: #1\par
     \end{indented}}

\begin{document}



\title{Deformation quantization and the tomographic representation of quantum fields}
\author{Jasel Berra--Montiel$^{1,2}$ and Roberto Cartas$^{3}$}

\address{$^{1}$ Facultad de Ciencias, Universidad Aut\'onoma de San Luis 
Potos\'{\i} 
Campus Pedregal, Av. Parque Chapultepec 1610, Col. Privadas del Pedregal, San
Luis Potos\'{\i}, SLP, 78217, Mexico}
\address{$^2$ Dual CP Institute of High Energy Physics, Mexico}
\address{$^3$ Instituto de F\'{i}sica, Universidad Aut\'onoma de Puebla,
Apartado postal J-48 72570 Puebla Pue., M\'exico.}

\eads{\mailto{\textcolor{blue}{jasel.berra@uaslp.mx}},\ 
\mailto{\textcolor{blue}{rcartas@ifuap.buap.mx}}\ 
}


\begin{abstract}
The tomographic representation of quantum fields within the deformation quantization formalism is constructed. By employing the Wigner functional we obtain the symplectic tomogram associated with quantum fields. In addition, the tomographic version of the Wigner map allows us to compute the symbols corresponding to field operators. Finally, the functional integral representation of the tomographic star product is determined. Some possible applications of the formalism to loop quantum cosmology and loop quantum gravity are briefly discussed. 
\end{abstract}

\keywords{Deformation quantization, star-product, tomographic representation}
\ams{81S30, 53D55, 70H45}


\section{Introduction}

Deformation quantization, also known in the literature as phase space quantum mechanics, consists of an alternative description of quantum mechanics based on the earlier works of Weyl \cite{Weyl}, Wigner \cite{Wigner}, Moyal \cite{Moyal}, Groenewold \cite{Groenewold} among others. A modern perspective of deformation quantization was later proposed by Bayen et al. in the seventies \cite{Bayen}, \cite{Flato}, and since then, many developments and applications beyond the earliest quantization problem, including areas with range in pure mathematics \cite{Kontsevich}, \cite{Waldmann}, quantum field theory \cite{Dito}, \cite{Compean}, \cite{Fredenhagen}, string theory \cite{Pinzul}, \cite{Bosonic} and loop quantum gravity \cite{DQpoly}, \cite{PolyWigner}, \cite{DQconstraints}, have been occurred. One of the main attributes of this formalism lies on a deformation of the algebra of observables with respect to some parameter (usually the Planck constant), so that, classical mechanics is fully recovered when the parameter is smoothly approached to zero. From the deformation quantization approach, the algebra of quantum observables is not given by a family of self-adjoint operators acting on a proper Hilbert space, as in the ordinary formulation of quantum mechanics. The quantum observables, instead, correspond to the set of smooth complex valued functions defined on the classical phase space, where the usual commutative point-wise product is replaced by a non-commutative product, known as the star product. In consequence, the non-commutative star product results in the deformation of the geometrical structures induced on the phase space, such as the Poisson algebra. Given that the employment of operators becomes unnecessary, within the deformation quantization scheme, an essential aspect relies on the definition of the Wigner distribution \cite{Wignerfunction}. This distribution stands for the symbol of the density matrix operator and contains the information related to all auto-correlation functions and transition amplitudes of a given quantum mechanical system. However, the Wigner distribution can not be interpreted as a probability distribution, as it usually acquires negative values on some domains along the phase space. Therefore, it
is usually referred instead as a quasi-probability distribution.

Recently, a new representation of quantum mechanics, called the tomographic or probability interpretation was proposed in \cite{Manko5}. This representation is based on the Radon transform of the Wigner distribution, and implements optical tomographic techiques, developed in quantum optics and quantum information, in order to reconstruct the Wigner distribution associated with the photon quantum states of the electromagnetic field \cite{Marmo}. In the tomographic representation, the quantum states are no longer described by wave functions but through tomographic probabilty distributions, also called marginal distributions, and the quantum evolution is dictated by a generalized Fokker-Planck type equation \cite{Fokker}. Over the last few years, this picture of quantum mechanics has been applied to problems related to quantum information \cite{QI}, statistical mechanics \cite{Statistics} and cosmology \cite{Cosmology} among others.

Following some ideas formulated in \cite{Manko8}, \cite{Manko3}, where the tomographic picture of the scalar field was obtained by considering the limit of an infinite set of tomograms associated with harmonic oscillators, the aim of this paper is to construct the tomographic representation by implementing the deformation quantization approach for general fields developed in \cite{Compean}, \cite{PolyWigner}. This approach, as we will demostrate below,  enable us to obtain the correspondence between the Wigner and the tomographic functionals for quantum fields. Moreover, by means of this correspondence, the integral representaion of the star product in the symplectic tomographic scheme is established. The formalism presented may be relevant in order to apply the tomographic representation to problems related to the semicalssical limit in loop quantum cosmology and loop quantum gravity, by using the methods formulated in \cite{DQpoly}, \cite{PolyWigner}.

The paper is organized as follows, in section 2, we briefly introduce the basic ideas of the deformation quantization scheme applied to fields. In section 3, the functional tomographic representation and the star product are obtained. Then, by employing the Wigner functional, the density operator in terms of the functional symplectic tomogram is established. To conclude, we introduce some final comments in section 4.         
 
\section{Deformation quantization of fields and the Wigner functional}

In this section, we introduce a concise review of the Wigner-Weyl quantization formalism for field theory. For this purpose, we will follow the formulation established in \cite{Compean} and \cite{PolyWigner}. For simplicity, throughout the manuscript we focus on the case of a real scalar field, nevertheless, a generalization to other class of fields can be derived straightforwardly. We also refer the interested reader to more detailed surveys on the deformation quantization \cite{Blaszak}, \cite{Bordemann}, \cite{Gutt}, \cite{Esposito}, \cite{Hirshfeld}, in order to get a more complete perspective of the quantization program.

Let us consider a real scalar field $\varphi$ defined on a 4D  Minkowski spacetime manifold $\mathcal{M}$. By performing a $3+1$ decomposition of the background spacetime in the form $\mathcal{M}=\Sigma\times\mathbb{R}$, for an arbitrary Cauchy surface $\Sigma$, which according to our current analysis, may be  considered as topologically equivalent to $\mathbb{R}^{3}$. The spacetime manifold $\mathcal{M}$ is endowed with a metric $\eta=diag(+1,+1.+1,-1)$ and local coordinates $(x,t)\in\mathbb{R}^{3}\times\mathbb{R}$. For the sake of simplicity, we deal with fields evaluated at the time instant $t=0$, and also we write $\varphi(x,0):=\varphi(x)$, and $\varpi(x,0):=\varpi(x)$, where $\varpi(x)$ stands for the canonically conjugate momentum variable associated with the field $\varphi(x)$. Therefore, the phase space of the field theory, denoted by $\Gamma$, is locally described in terms of coordinates $(\varphi,\varpi)$, which in turn provide a specific set of initial conditions on the Cauchy surface $\Gamma$ \cite{Witten}.

According to the axioms of quantum mechanics, in order to define a quantization rule \cite{Weyl}, that is, a one-to-one mapping that associates to each real-valued function $F$ on the classical phase space $\Gamma$, a self-adjoint operator $\hat{F}$ on the quantum Hilbert space $\mathcal{H}$; it is necessary to provide a quantization map such that it takes the canonical Poisson brackets
\begin{eqnarray}
\left\lbrace \varphi(x),\varpi(y)\right\rbrace=\delta(x-y),\nonumber \\
\left\lbrace \varphi(x),\varphi(y)\right\rbrace=0=\left\lbrace \varpi(x),\varpi(y)\right\rbrace,
\end{eqnarray}
to the Heisenberg commutation relations between self-adjoint operators (in natural units where $\hbar=1$)
\begin{eqnarray}
\left[\hat{\varphi}(x),\hat{\varpi}(y)\right]\Psi=i\delta(x-y)\Psi,\nonumber\\
\left[\hat{\varphi}(x),\hat{\varphi}(y)\right]\Psi=0=\left[\hat{\varpi}(x),\hat{\varpi}(y)\right]\Psi,
\end{eqnarray}
in the case of a scalar field theory, the vector state $\Psi[\varphi]$ belongs to the Hilbert space $\mathcal{H}=L^{2}(\mathcal{S}'(\mathbb{R}^{3}),d\mu)$. To be more specific, $\mathcal{S}'(\mathbb{R}^{3})$  represents the Schwartz space of tempered distributions, that is, the space of continuous linear functionals of rapidly decreasing smooth test functions $\mathcal{S}(\mathbb{R}^{3})$, and $d\mu$ symbolizes the formal measure $\mathcal{D}\varphi:=\prod_{x\in\mathbb{R}^{3}}d\varphi(x)$, which represents a formal uniform Lebesgue measure on the field configuration space \cite{Glimm}.  

Given $F[\varphi,\varpi]$ a functional on the phase space $\Gamma$, let us denote by $\tilde{F}[\lambda,\mu]$ its corresponding Fourier transform, 
\begin{equation}\label{Fourier}
\tilde{F}[\lambda,\mu]=\int\mathcal{D}\varphi\mathcal{D}\varpi\exp\left\lbrace-i\int_{\mathbb{R}^{3}} dx(\lambda(x)\varphi(x)+\mu(x)\varpi(x)) \right\rbrace F[\varphi,\varpi], 
\end{equation}
where $\lambda(x),\mu(x)\in\mathcal{S}(\mathbb{R}^{3})$. Then, analogously to quantum mechanics, implementing the Weyl-Wigner quantization process for fields means to define a linear map, $W:L^{2}(\Gamma)\to\mathcal{L}(\mathcal{H})$, from the space of functionals on the phase space $\Gamma$, into the space of linear operators on the Hilbert space $\mathcal{H}$ (with the symmetric ordering of the position and momentum operators) \cite{Compean}, \cite{Stratonovich}, given by 
\begin{equation}\label{Wrule}
\hat{F}\Psi=W(F[\varphi,\varpi])\Psi:=\int\mathcal{D}\varphi\mathcal{D}\left(\frac{\varpi}{2\pi}\right) \tilde{F}[\lambda,\mu]\hat{\mathcal{U}}[\lambda,\mu], 
\end{equation}
where $\hat{\mathcal{U}}[\lambda,\mu]$ corresponds to a family of unitary operators on $\mathcal{H}$ defined as
\begin{equation}
\hat{\mathcal{U}}[\lambda,\mu]=\exp\left\lbrace i\int_{\mathbb{R}^{3}} dx\,(\lambda(x)\hat{\varphi}(x)+\mu(x)\hat{\varpi}(x))\right\rbrace. 
\end{equation}
Employing the completness relations $\int\mathcal{D}\varphi\ket{\varphi}\bra{\varphi}=1$ and $\int\mathcal{D}\left( \frac{\varpi}{2\pi}\right) \ket{\varpi}\bra{\varpi}=1$, where $\ket{\varphi}$ and $\ket{\varpi}$ stand for vector eigenstates of the operators $\hat{\varphi}$ and $\hat{\varpi}$ respectively, such that, $\hat{\varphi}(x)\ket{\varphi}=\varphi(x)\ket{\varphi}$, and $\hat{\varpi}(x)\ket{\varpi}=\varpi(x)\ket{\varpi}$, it can be shown that the operator $\hat{\mathcal{U}}[\lambda,\mu]$ satisfies the following properties \cite{Zachos}:
\begin{eqnarray}\label{trace1}
\tr\left\lbrace \hat{\mathcal{U}}[\lambda,\mu]\right\rbrace =\int\mathcal{D}\varphi\braket{\varphi|\,\hat{\mathcal{U}}[\lambda,\mu]\,|\varphi}=\delta\left[\frac{\lambda}{2\pi}\right] \delta\left[\mu\right], \\
\tr\left\lbrace \hat{\mathcal{U}}^{\dagger}[\lambda,\mu]\hat{\mathcal{U}}[\lambda',\mu']\right\rbrace=\delta\left[\frac{\lambda-\lambda'}{2\pi}\right] \delta\left[\mu-\mu'\right].\label{trace2}
\end{eqnarray} 

By inserting the Fourier transform (\ref{Fourier}) into the expression (\ref{Wrule}) one obtains
\begin{equation}\label{quantizer}
\hat{F}\Psi=W(F[\varphi,\varpi])\Psi=\int\mathcal{D}\varphi\mathcal{D}\left( \frac{\varpi}{2\pi}\right)F[\varphi,\varpi]\hat{\Delta}[\varphi,\varpi]\Psi, 
\end{equation} 
where the operator $\hat{\Delta}[\varphi,\varpi]$ reads
\begin{equation}\label{Stratonovich}
\mkern-20mu\hat{\Delta}[\varphi,\varpi]=\int\mathcal{D}\left(\frac{\lambda}{2\pi}\right)\mathcal{D}\mu \exp\left\lbrace -i\int_{\mathbb{R}^{3}}dx(\lambda(x)\varphi(x)+\mu(x)\varpi(x))\right\rbrace\hat{\mathcal{U}}[\lambda,\mu]. 
\end{equation}
The operator $\hat{\Delta}[\varphi,\varpi]$ represented in (\ref{Stratonovich}) corresponds to the quantum field analog of the Weyl-Stratonovich quantizer \cite{Stratonovich}, and as we shall observe in brief, it not only allow us additionally to define the symbol associated with a quantum operator, but to construct an associative and noncommutative product between classical observables, the so called star-product. By using the properties (\ref{trace1}) and (\ref{trace2}), the operator $\hat{\Delta}[\varphi,\varpi]$ fulfills the subsequent relations 
\begin{eqnarray}
\hat{\Delta}^{\dagger}[\varphi,\varpi]=\hat{\Delta}[\varphi,\varpi],\\
\tr\left\lbrace \hat{\Delta}[\varphi,\varpi]\right\rbrace =1, \\
\tr\left\lbrace \hat{\Delta}[\varphi,\varpi]\hat{\Delta}[\varphi',\varpi']\right\rbrace=\delta[\varphi-\varphi']\delta\left[ \frac{\varpi-\varpi'}{2\pi}\right].\label{compatibility} 
\end{eqnarray}
The expression depicted in (\ref{compatibility}) is known as the compatibility condition \cite{Manko1}, and enables us to obtain the phase space functional associated with a quantum field operator by means of the so-called Wigner map
\begin{equation}\label{dequantizer}
F[\varphi,\varpi]=W^{-1}(\hat{F})=\tr\left\lbrace \hat{\Delta}[\varphi,\varpi]\hat{F}\right\rbrace. 
\end{equation}
This map corresponds to the inverse relation of the Weyl-Stratonovich quantizer (\ref{quantizer}), and it associates quantum operators to real functions, also referred as symbols, following the common terminology used in harmonic analysis \cite{Reed}. This means that the operator $\hat{\Delta}[\varphi,\varpi]$, through the application of the trace given in (\ref{dequantizer}), allows us to dequantize quantum observable $\hat{F}$ \cite{Manko1}. The symbols obtained by means of the Wigner map determine an associative algebra endowed with a noncommutative product denominated as the Moyal star-product for field functionals. Let $F_{1}[\varphi,\varpi]=W^{-1}(\hat{F}_{1})$ and $F_{2}[\varphi,\varpi]=W^{-1}(\hat{F}_{2})$ field functionals, the symbol associated with the product of the following operators $\hat{F}_{1}\hat{F}_{2}$ becomes
\begin{equation}\label{Moyal}
(F_{1}\star F_{2})[\varphi,\varpi]=W^{-1}(\hat{F}_{1}\hat{F}_{2})=\tr\left\lbrace \hat{\Delta}[\varphi,\varpi]\hat{F}_{1}\hat{F}_{2}\right\rbrace.
\end{equation}
By substituting the Weyl-Stratonovich operator $\hat{\Delta}[\varphi,\varpi]$ into (\ref{Moyal}) and using the completeness relations for the eigenstates $\ket{\varphi}$ (see \cite{Compean}, \cite{HirshfeldP} for further details), one obtains
\begin{equation}
\mkern-110mu(F_{1}\star F_{2})[\varphi,\varpi]=\int\mathcal{D}\varphi'\mathcal{D}\varphi''\mathcal{D}\left(\frac{\varpi'}{\pi}\right) \mathcal{D}\left(\frac{\varpi''}{\pi} \right)F_{1}[\varphi',\varpi']F_{2}[\varphi'',\varpi'']K[\varphi,\varpi,\varphi',\varpi',\varphi'',\varpi''], 
\end{equation}
where the integral Kernel $K[\varphi,\varpi,\varphi',\varpi',\varphi'',\varpi'']$ takes the form
\begin{equation}
\mkern-100mu K[\varphi,\varpi,\varphi',\varpi',\varphi'',\varpi'']=\exp\left\lbrace 2i\int_{\mathbb{R}^{3}}dx\left((\varphi-\varphi')(\varpi-\varpi'')-(\varphi-\varphi'')(\varpi-\varpi') \right) \right\rbrace.
\end{equation}

Let us now finish this section by defining the quantum field analog of the Wigner distribution corresponding to the density operator $\hat{\rho}=\ket{\Psi}\bra{\Psi}$ of a given quantum state $\Psi\in\mathcal{H}$ \cite{Zachos}. By means of the Wigner map (\ref{dequantizer}), the symbol associated with the operator $\hat{\rho}$ satisfies
\begin{eqnarray}\label{rho}
\rho[\varphi,\varpi]=\tr\left\lbrace \hat{\Delta}\hat{\rho}\right\rbrace, \nonumber \\
=\int\mathcal{D}\left( \frac{\xi}{2\pi}\right)\exp\left\lbrace -i\int_{\mathbb{R}^{3}}dx\,\xi(x)\varpi(x)\right\rbrace \Psi^{*}\left[\varphi-\frac{\xi}{2}\right]\Psi\left[\varphi+\frac{\xi}{2}\right].    
\end{eqnarray}
A notable property for this functional lies on the possibility to take negative values on some regions along the phase space, therefore, it can not be interpreted as a probability distribution, thus it
is usually referred as a quasi-probability distribution. This distinctive feature makes the Wigner distribution an excellent indicative of quantum properties, since, precicely the negative contributions tend to be related to nonclassical interferences, although not all quantum states have negative contributions to the Wigner distribution \cite{Curtright}.

\section{The tomographic representation of quantum fields}

In this section, we derive the tomographic representation of quantum fields by employing the deformation quantization formalism and the Wigner functional presented in the previous section. We will first analyze the symplectic tomograms by means of the Wigner functional, and then, with the aid of the quantum field analog of the Weyl-Stratonovich tomogram operator, we will obtain the corresponding star product between functionals.

\subsection{The symplectic tomogram}

As a first step, let us consider the field quadrature associated with the quantum field observable $\hat{\chi}(x)$, which corresponds to a generic linear combination of the field operators $\hat{\varphi}$ and $\hat{\varpi}$ of the form
\begin{equation}
\hat{\chi}(x)=\mu(x)\hat{\varphi}(x)+\nu(x)\hat{\varpi}(x),
\end{equation}
where $\mu(x), \nu(x)$ are real functions labelling different reference frames in the phase space  \cite{Manko3}, \cite{Manko2}. Given the density operator $\hat{\rho}$ of a quantum system, let us define the characteristic functional $G[\lambda]$, as the mean value of the exponential of the operator $\hat{\chi}(x)$
\begin{equation}\label{G}
G[\lambda]:=\left\langle \exp\left\lbrace i\int_{\mathbb{R}^{3}}dx\lambda(x)\hat{\chi}(x) \right\rbrace \right\rangle=\tr\left\lbrace  \hat{\rho}\exp\left(  i\int_{\mathbb{R}^{3}}dx\,\lambda(x)\hat{\chi}(x)\right)  \right\rbrace.
\end{equation} 
Then, by applying the Fourier transform of the functional $G[\lambda]$, we obtain
\begin{equation}\label{tomogram}
w[\chi,\mu,\nu]:=\int\mathcal{D}\left(\frac{\lambda}{2\pi}\right) G[\lambda]\exp\left\lbrace -i\int_{\mathbb{R}^{3}}dx\,\lambda(x)\chi(x)\right\rbrace.  
\end{equation}
The functional $w[\chi,\mu,\nu]$ corresponds to the quantum field analog of the marginal distribution function established in \cite{Manko5}, \cite{Manko4}, and as we will observe, it completely determines a quantum system by describing quantum field states as genuine probability distribution functionals. By substituting $G[\lambda]$ into the expression (\ref{tomogram}), and writing the trace of the density operator $\hat{\rho}$ in terms of the Wigner functional $\rho[\varphi,\varpi]$, as stated in (\ref{rho}), we obtain that $w[\chi,\mu,\nu]$ holds
\begin{eqnarray}\label{tomogramW}
w[\chi,\mu,\nu]&=&\int\mathcal{D}\left(\frac{\lambda}{2\pi}\right)\mathcal{D}\varphi\mathcal{D}\left(\frac{\varpi}{2\pi}\right)\rho[\varphi,\varpi] \nonumber\\
&&\times \exp\left\lbrace -i\int_{\mathbb{R}^{3}}dx\,\lambda(x)(\chi(x)-\mu(x)\varphi(x)-\nu(x)\varpi(x)) \right\rbrace.   
\end{eqnarray}
In analogy with the case of finite degrees of freedom \cite{Manko5}, we call the functional $w[\chi,\mu,\nu]$ the symplectic tomogram of quantum fields. By evaluating the trace in (\ref{G}) on eignestates of the operator $\hat{\chi}(x)$, that is $\hat{\chi}(x)\ket{\chi}=\chi(x)\ket{\chi}$, it can be verified that  the symplectic tomogram $w[\chi,\mu,\nu]$ is proportional to $\braket{\chi|\,\hat{\rho}\,|\chi}$, which means that, as a functional, it is positive and normalized, yielding
\begin{equation}
\int\mathcal{D}\chi\, w[\chi,\mu,\nu]=1.
\end{equation}   
The formula (\ref{tomogramW}) can be inverted, which allows us to express the Wigner functional in terms of the symplectic tomogram as
\begin{equation}\label{stomogram}
\mkern-100mu \rho[\varphi,\varpi]=\int\mathcal{D}\chi\mathcal{D}\mu\mathcal{D}\left(\frac{\nu}{2\pi}\right)\,w[\chi,\mu,\nu]\exp\left\lbrace -i\int_{\mathbb{R}^{3}}(\mu(x)\varphi(x)+\nu(x)\varpi(x)-\chi(x)) \right\rbrace.  
\end{equation}
Considering that the Wigner functional, $\rho[\varphi,\varpi]$ as a quasi-probability distribution, suffices to determine entirely the quantum state of a system, and in addition, the Wigner functional is in turn completely formulated in terms of the symplectic tomogram (which results to be a positive and normalized distribution), this means that the information  associated with a quantum field state is completely encoded on actual probaility distributions, just as in the case of classical mechanics \cite{Manko5}. This representation of quantum mechanics, usually referred as the probability representation, was first introduced in \cite{Manko4}, \cite{Manko6}, to describe systems with finite degrees of freedom.    

If we express the density operator $\hat{\rho}$ in terms of the Wigner functional $\rho[\varphi,\varpi]$ by means of the formula (\ref{quantizer}), as
\begin{equation}
\hat{\rho}=\int\mathcal{D}\varphi\mathcal{D}\left(\frac{\varpi}{2\pi}\right)\rho[\varphi,\varpi]\hat{\Delta}[\varphi,\varpi],
\end{equation}
and replacing the Wigner functional by the equation (\ref{stomogram}), then the density operator reads
\begin{equation}\label{tdensity}
\mkern-90mu\hat{\rho}=\int\mathcal{D}\chi\mathcal{D}\mu\mathcal{D}\left(\frac{\nu}{2\pi} \right)w[\chi,\mu,\nu]\exp\left\lbrace -i\int_{\mathbb{R}^{3}}dx\,(\chi(x)-\mu(x)\hat{\varphi}(x)-\nu(x)\hat{\varpi}(x))\right\rbrace,
\end{equation}  
which shows that the density operator for quantum fields, analogously to the case of quantum mechanics \cite{Manko2}, can be reconstructed by using the information carried by the symplectic tomogram.

\subsection{The star product}

Within the symplectic tomographic scheme for quantum fields, and by using the approach developed in \cite{Manko7}, the tomographic symbol associated with a field operator $\hat{F}\in\mathcal{L}(\mathcal{H})$, is obtained by means of the operator
\begin{equation}\label{tomsymbol}
\hat{\Delta}^{T}[\chi, \mu,\nu]:=\delta[\chi(x)\hat{1}-\mu(x)\hat{\varphi}(x)-\nu(x)\hat{\varpi}(x)],
\end{equation}
where $\hat{1}$ stands for the identity operator. This means that
\begin{equation}\label{tdequantizer}
F[\chi,\mu,\nu]:=W^{-1}_{T}(\hat{F})=\tr\left\lbrace\hat{\Delta}^{T}[\chi,\mu,\nu]\hat{F} \right\rbrace,
\end{equation}
being $W^{-1}_{T}$ the tomographic version of the Wigner map (\ref{dequantizer}). The corresponding compatibility condition in the tomographic representation, analogous to the case of the Weyl-Stratonovich operator presented in (\ref{compatibility}), is given by
\begin{equation}\label{tcompatibility}
\tr\left\lbrace \hat{\Delta}^{T}[\chi,\mu,\nu]\hat{\mathcal{C}}[\chi',\mu',\nu'] \right\rbrace=\delta[\chi-\chi']\delta[\mu-\mu']\delta\left[\frac{\nu-\nu'}{2\pi} \right],   
\end{equation}
where the operator $\hat{\mathcal{C}}[\chi,\mu,\nu]$ or dequantizer reads
\begin{equation}
\hat{\mathcal{C}}[\chi,\mu,\nu]:=\exp\left\lbrace i\int_{\mathbb{R}^{3}}dx\,(\chi(x)\hat{1}-\nu(x)\hat{\varpi}(x)-\mu(x)\hat{\varphi}(x))\right\rbrace. 
\end{equation}
By using equation (\ref{tcompatibility}), we can invert the relation (\ref{tdequantizer}), so that
\begin{equation}\label{tquantizer}
\hat{F}:=W_{T}(F)=\int\mathcal{D}\chi\mathcal{D}\mu\mathcal{D}\left(\frac{\nu}{2\pi} \right)F[\chi,\mu,\nu]\hat{\mathcal{C}}[\chi,\mu,\nu].
\end{equation}
Making use of the tomographic version of the Wigner map (\ref{tdequantizer}), let us compute the symbols associated with the canonical variables in field theory,
\begin{eqnarray}
\varphi[\chi,\mu,\nu]&:=&W^{-1}_{T}(\hat{\varphi})=\tr\left\lbrace \hat{\Delta}^{T}\hat{\varphi}\right\rbrace, \nonumber \\ &=&i\left( \frac{\delta}{\delta \mu(x)} \delta[\mu]\right) \delta[\nu]\int\mathcal{D}\lambda\frac{1}{\lambda^{2}(x)}\exp\left\lbrace -i\int_{\mathbb{R}^{3}}\lambda(x)\chi(x)\right\rbrace, \nonumber \\
&=& i \left( \frac{\delta}{\delta \mu(x)} \delta[ \mu]\right)\delta\left[ \frac{2\nu}{\pi}\right]\chi(x)|\chi(x)|,  
\end{eqnarray}
and similarly
\begin{eqnarray}
\varpi[\chi,\mu,\nu]&:=&W^{-1}_{T}(\hat{\varpi})=\tr\left\lbrace \hat{\Delta}^{T}\hat{\varpi}\right\rbrace, \nonumber \\ 
&=& i \left( \frac{\delta}{\delta\nu(x)} \delta[ \nu]\right)\delta\left[ \frac{2\mu}{\pi}\right]\chi(x)|\chi(x)|,  
\end{eqnarray}
The symbols $\varphi[\chi,\mu,\nu]$ and $\varpi[\chi,\mu,\nu]$, extend the notion of the quantum tomographic symbols related to the position and momentum operators obtained in \cite{Pilyavets}, to systems with an infinite number of degrees of freedom.

Now, with the compatibility condition at hand (\ref{tcompatibility}), it is possible to obtain the star product associated with the product of a pair of field functionals in the tomographic representation. Let $F_{1}[\chi,\mu,\nu]$ and $F_{2}[\chi,\mu,\nu]$ the symbols for the operators $\hat{F}_{1}$, $\hat{F}_{2}\in\mathcal{L}(\mathcal{H})$ in the tomographic scheme, respectively. The tomographic symbol corresponding to the product of the operators $\hat{F}_{1}\hat{F}_{2}$ becomes
\begin{equation}
(F_{1}\star F_{2})[\chi,\mu,\nu]=\tr\left\lbrace \hat{\Delta}^{T}[\chi,\mu,\nu]\hat{F}_{1}\hat{F}_{2}\right\rbrace. 
\end{equation}
By writing $\hat{F}_{1}$ and $\hat{F}_{2}$ as in the formula (\ref{tquantizer}), and making use of the compatibility relation (\ref{tcompatibility}), after several simple functional integrations one can verify that
\begin{eqnarray}\label{tstar}
\mkern-100mu(F_{1}\star F_{2})[\chi,\mu,\nu]&=&\int \mathcal{D}\chi'\mathcal{D}\mu'\mathcal{D}\left(\frac{\nu'}{2\pi} \right)\mathcal{D}\chi''\mathcal{D}\mu''\mathcal{D}\left(\frac{\nu''}{2\pi} \right) F_{1}[\chi',\mu',\nu']F_{2}[\chi'',\mu'',\nu''] \nonumber\\
&& \times K[\chi,\mu,\nu, \chi',\mu',\nu', \chi'',\mu'',\nu''], 
\end{eqnarray}
where the integral kernel $K[\chi,\mu,\nu, \chi',\mu',\nu', \chi'',\mu'',\nu'']$ takes the form
\begin{eqnarray}
\mkern-100mu K[\chi,\mu,\nu, \chi',\mu',\nu', \chi'',\mu'',\nu'']=\delta[\mu(\nu'+\nu'')-\nu(\mu'+\mu'')]\\
\mkern-100mu \times \exp\left\lbrace \frac{i}{2}\int_{\mathbb{R}^{3}}dx\,\left( \nu'(x)\mu''(x)-\nu''(x)\mu'(x)+2\chi'(x)+2\chi''(x)-\frac{2\chi(x)}{\nu(x)}(\nu'(x)+\nu''(x))\right)  \right\rbrace. \nonumber
\end{eqnarray}
The star product (\ref{tstar}), obtained by means of the functional representation, agrees with the tomographic star product computed for the scalar field \cite{Manko8}, where it has been considered as a starting point the expansion of the fields $\varphi$ and $\varpi$ as described by an infinite set of harmonic oscillators.

To finish this section, it is worthwhile to mention that we can make use the symplectic tomogram $w[\chi,\mu,\nu]$, in order to compute the expectation value of any quantum field observable $\hat{F}\in\mathcal{L}(\mathcal{H})$. By employing the density operator in terms of the tomogram (\ref{tdensity}) and the relation (\ref{tquantizer}) between operators and their corresponding symbols, we obtain
\begin{equation}
\braket{\hat{F}}=\tr\left\lbrace \hat{\rho}\hat{F} \right\rbrace= \int\mathcal{D}\chi\mathcal{D}\mu\mathcal{D}\left(\frac{\nu}{2\pi}\right)\,w[\chi,\mu,\nu]\,\tr\left\lbrace\hat{F}\,\hat{\mathcal{C}}[\chi,\mu,\nu]\right\rbrace. 
\end{equation} 
In this sense, expectation values of physical observables are computed through the integration with the functional symplectic tomogram, in close analogy with the Wigner distribution functional, except for the positivity definiteness of the tomogram, which provides a genuine probability distribution. 

\section{Conclusions}
\label{sec:conclu}  
 
In this paper, we have discussed some aspects of the tomographic representation for quantum fields within the deformation quantization formalism. We first introduced a brief review of the Wigner-Weyl quantization  procedure for fields, emphasazing the role performed by the star product on the algebra of quantum observables. Further, within our context, the tomographic representation was derived by employing the Wigner functional corresponding to the density operator of a quantum field state. In this manner, we showed that the information related to a quantum field is entirely encoded in terms of the functional symplectic tomogram, which in turn, provides a genuine probability distribution, contrary to the case of the Wigner functional. Moreover, by making use of the tomographic version of the Wigner map, the symbols associated with the canonical field operators have been computed. Finally, we have determined the integral representation of the tomographic star product for any pair of functionals defined on the phase space. We expect that the tomographic representaton developed here may be relevant in order investigate aspects related to the semiclassical limit of theories in the framework of loop quantum gravity and loop quantum cosmology. According to loop quantum gravity, matter fields can only be supported on spin-networks, i.e., polymer-like excitations of quantum geometry states. Within this formulation, the algebra of observables for a scalar field is generated by the space of cylindrical functions supported on spin-networks, also known as scalar network functions. Indeed, by applying the methods developed in \cite{DQpoly}, \cite{PolyWigner}, it has been proved that the Wigner distribution associated with a real-valued scalar field in the loop representation, is given by the limit of Gaussian measures in the Schr\"odinger functional representation. This limit, by means of the Bochner-Minlo's theorem, is defined on the space of generalized distributions over the Bohr compactification of the reals. The aforementioned analysis suggests that a similar construction may be implemented for the tomographic distribution. Indeed, since the symplectic tomogram of a quantum field is essentially the Radon transform of the Wigner functional, it is compelling to determine if the former limit preserves the positivity of the tomographic picture over the Bohr compactification.   
Furthermore, considering that within the tomographic scheme, classical and quantum states are depicted by the same objects, that is, positive probability distributions, this distinctive characteristic could pave the way in order to address the problem of study semiclassical limits associated with results originated from models in loop quantum cosmology with a different perspective. We intend to dedicate a future publication to address these issues.

\section*{Acknowledgments}
The authors would like to acknowledge financial support from CONACYT-Mexico
under project CB-2017-283838.

\section*{References}

\bibliographystyle{unsrt}

\begin{thebibliography}{l}
\bibitem{Weyl}Weyl H., \emph{Group Theory and Quantum Mechanics}, (Dover, New York, 1931). 

\bibitem{Wigner}Wigner E., \emph{On the Quantum Correction For Thermodynamic Equilibrium}, Phys.~Rev.~{\bf 40} 749 (1932).

\bibitem{Moyal}Moyal J.~E., \emph{Quantum mechanics as a statistical theory}, Poc.~Cambridge Philos.~Soc. {\bf 45} 99-124 (1949).

\bibitem{Groenewold}Groenewold H.~J, \emph{On the principles of elementary quantum mechanics}, Physica (Utrecht) {\bf 12} 405-460 (1946).

\bibitem{Bayen}F.~Bayen, M.~Flato, C.~Fronsdal, A.~Lichnerowicz and D.~Sternheimer, \emph{Deformation theory and quantization I. Deformations of symplectic structures}, Ann.~Phys.~{\bf 111}, 61--110 (1978).

\bibitem{Flato}F.~Bayen, M.~Flato, C.~Fronsdal, A.~Lichnerowicz and D.~Sternheimer, \emph{Deformation theory and quantization. II. Physical applications}, Ann.~Phys.~{\bf 111}, 111--51 (1978).

\bibitem{Kontsevich}Kontsevich M., \emph{Deformation quantization of Poisson manifolds}, Lett.~Math.~Phys.~{\bf 66} 157–216 (2003), \texttt{arXiv:q-alg/9709040}. 

\bibitem{Waldmann}Waldmann S-, \emph{Recent Developments in Deformation Quantization} ed. F. Finster et al. (Quantum Mathematical Physics) (Basel: Birkhauser, 2016), \texttt{arXiv:1502.00097 [math.QA]}.

\bibitem{Dito}Dito J., \emph{Star-product approach to quantum field theory: The free scalar field}, Lett.~Math.~Phys.~{\bf 20 } 125-134 (1990).

\bibitem{Compean} Garc\'ia-Compean H., Plebansky J.~F., Przanowski M., and Turrubiates F.~J., \emph{Deformation quantization of classical fields}, Int.~J.~Mod.~Phys.~{\bf A 16} 2533-58 (2001), \texttt{arXiv:hep-th/9909206}.

\bibitem{Fredenhagen}Fredenhagen K. and Rejzner K., \emph{QFT on curved spacetimes: axiomatic framework and examples}, J.~Math.~Phys.~{\bf 57} 031101 (2016), \texttt{arXiv:1412.5125 [math-ph]}.


\bibitem{Pinzul}Pinzul A. and Stern A., \emph{Absence of the holographic principle in noncommutative Chern-Simons theory}, JHEP~{\bf 11} 023 (2011), \texttt{arXiv:hep-th/0107179}.

\bibitem{Bosonic}Gracia-Compean H., Plebanski J.~F., Przanowski M. and Turrubiates J.~F., \emph{Deformation Quantization of Bosonic Strings}, J.~Phys.~A {\bf 33} 7935-7954 (2000), \texttt{arXiv:hep-th/0002212}.


\bibitem{DQpoly}Berra-Montiel J. and Molgado A., \emph{Polymer quantum mechanics as a deformation quantization}, Class.~Quantum Grav.~{\bf 36} 025001 (2019), \texttt{arXiv:1805.05943 [gr-qc]}.

\bibitem{PolyWigner}Berra-Montiel J., \emph{The Polymer representation for the scalar field: a Wigner functional approach}, Class.~Quantum Grav.~{\bf 37} 025006 (2020), \texttt{arXiv:1908.09194 [gr-qc]}.

\bibitem{DQconstraints}Berra-Montiel J. and Molgado A., \emph{Deformation quantization of constrained
systems: a group averaging approach}, Class.~Quantum Grav.~{\bf 37} 055009 (2020), \texttt{arXiv: 1911.00945 [gr-qc]}.

\bibitem{Wignerfunction}Wigner E., \emph{Do the Equations of Motion Determine the Quantum Mechanical Commutation Relations?}, Phys.~Rev.~{\bf 77} 711 (1950).

\bibitem{Manko5}Mancini S., Man'ko V.~I. and Tombesi P., \emph{Symplectic tomography as classical approach
to quantum systems}, Phys.~Lett.~A~{\bf 213} 1-6 (1996), \texttt{arXiv:quant-ph/9603002}.

\bibitem{Marmo}Ibort A., Man'ko V.~I., Marmo G., Simoni A. and Ventriglia F., \emph{An introduction to the tomographic picture of quantum mechanics}, Phys.~Scr.~{\bf 79} 065013 (2009), \texttt{arXiv:0904.4439 [quant-ph]}.

\bibitem{Fokker}Mancini S., Man'ko V.~I and Tombesi P., \emph{Classical-Like Description of Quantum Dynamics by Means of Symplectic Tomography}, Found.~Phys.~{\bf 27} 801 (1997), \texttt{arXiv:quant-ph/9609026}.

\bibitem{QI}Helsen J., Battistel J. and Terhal B.~M., \emph{Spectral quantum tomography}, npj Quantum Inf.~{\bf 5} 74 (2019), \texttt{arXiv:1904.00177 [quant-ph]}.

\bibitem{Statistics}Man'ko V.~I. and Mendes R.~V., \emph{Lyapunov Exponent In Quantum Mechanics. A Phase-space Approach}, Physica D~{\bf 145} 33-348 (2000), \texttt{arXiv:quant-ph/0002049}.

\bibitem{Cosmology}Capozziello S., Man'ko V.~I., Marmo G. and Stornaiolo C., \emph{A tomographic description for classical and quantum cosmological perturbations}, Phys.~Scr.~{\bf 80} 045901 (2009), \texttt{arXiv:0905.1244 [gr-qc]}.

\bibitem{Manko8}Man'ko M.~A., Man'ko V.~I. and Thanh N.~C., \emph{ Tomographic-probability representation of the quantum scalar field}, J.~Russ.~Laser Res.~{\bf 30} 1-11 (2009).

\bibitem{Manko3}Man'ko V.~I., Rosa L. and Vitale P., \emph{Probability representation in quantum field theory}, Phys.~Lett.~B~{\bf 439} 328-336 (1998), \texttt{arXiv:hep-th/9806164}.






\bibitem{Blaszak}Blaszak M. and Domanski Z., \emph{Phase space quantum mechanics} Ann.~Phys.~{\bf 327} 167-211 (2012), \texttt{arXiv:1009.0150 [math-ph]}.

\bibitem{Bordemann}Bordemann M., \emph{Deformation quantization: a survey}, J.~Phys.:~Conf.~Ser.~{\bf 103} 012002 (2008).

\bibitem{Gutt}Gutt S., \emph{Deformation quantisation of Poisson manifolds}, Geom.~Topol.~Monogr.~{\bf 17} 171-220 (2001).

\bibitem{Esposito}Esposito C., \emph{Formality Theory: From Poisson Structures to Deformation Quantization}, (New York, Springer, 2015).

\bibitem{Hirshfeld}Hirshfeld A.~C and Henselder P., \emph{Deformation quantization in the teaching of quantum mechanics}, Am.~J.~Phys.~{\bf 70} 537 (2002), \texttt{arXiv:quant-ph/0208163}.

\bibitem{Witten}Crnkovic C. and Witten E., \emph{Covariant Description of Canonical Formalism in Geometrical Theories}, in Hawking S.~W. and Israel W. (ed.): Three hundreds years of gravitation, 676-684 (Princeton, 2986). 



\bibitem{Glimm} Glimm J. and Jaffe A., \emph{Quantum Physics, a Functional Integral Point of View} (Berlin: Springer, 1987).

\bibitem{Stratonovich}Stratonovich R.~L., \emph{On the statistical interpretation of quantum theory} Sov.~Phys.~JETP {\bf 31}, 1012 (1956).

\bibitem{Zachos}Curtright T. and Zachos C., \emph{Wigner Trajectory Characteristics in Phase Space and Field Theory}, J.~Phys.~A {\bf 32} 771-779 (1999), \texttt{arXiv:hep-th/9810164}.

\bibitem{Manko1}Man'ko O.~V., Man'ko V.~I., Marmo G. and Vitale P., \emph{Star products, duality and double Lie algebras}, Phys.~Lett.~A~{\bf 360} 522-532 (2007), \texttt{arXiv:quant-ph/0609041}.

\bibitem{Reed}Reed M. and Simon B., \emph{Methods of Modern Mathematical Physics}, Vol II (United States Academic Press, 1975).

\bibitem{HirshfeldP}Hirshfeld A.~C. and Henselder P., \emph{Star Products and Perturbative Quantum Field Theory}, Ann.~Phys. {\bf 298} 382-393 (2002), \texttt{arXiv:hep-th/0208194}.

\bibitem{Curtright}Zachos C.~K., Fairlie D.~B. and Curtright T.~L., \emph{Quantum Mechanics ins Phase Space: An Overview with Selected Papers}, (Singapure, World-Scientific, 2005).

\bibitem{Manko2}D'Ariano G.~M., Mancini S., Man'ko V.~I. and Tombesi P., \emph{Reconstructing the density operator by using generalized field quadratures}, Quantum Semiclass.~Opt.~{\bf 8} 1017-1027 (1996), \texttt{arXiv:quant-ph/9606034}.
 

\bibitem{Manko4}Mancini S., Man'ko V.~I. and Tombesi P., \emph{Wigner function and probability distribution for shifted and squeezed quadratures}, Quantum Semiclass.~Opt.~{\bf 7} 615-623 (1995).


\bibitem{Manko6}Mancini S., Tombesi P. and Man'ko V.~I., \emph{Density Matrix From Photon Number Tomography}, EPL~{\bf 37} 79 (1997), \texttt{arXiv:quant-ph/9612004}.

\bibitem{Manko7}Man'ko M.~A., Man'ko V.I, and Vilela-Mendes R., \emph{Tomograms and other transforms: a unified view}, J.~Phys.~A: Math.~Gen.~{\bf 34} 8321-8332 (2001), \texttt{arXiv:math-ph/0101025}.

\bibitem{Pilyavets}Man'ko M.~A., Man'ko V.I and Pilyavets O.~V., \emph{Probability representation of classical states}, J.~Russ.~Laser Res.~{\bf 26} 429-444 (2005).







\end{thebibliography}

\end{document}